\documentclass[
reprint,
superscriptaddress,
nofootinbib,
amsmath,
amssymb,
aps,
pra,
floatfix,
longbibliography,
]{revtex4-2}

\usepackage{pdfpages}

\makeatletter
\AtBeginDocument{\let\LS@rot\@undefined}
\makeatother

\usepackage{pgffor}

\usepackage[utf8]{inputenc}
\DeclareUnicodeCharacter{00A0}{~} 

\usepackage{siunitx}
\usepackage{graphicx}

\usepackage{dcolumn}
\usepackage{bm}
\usepackage{comment}
\usepackage{chemformula}
\usepackage{lineno}

\preprint{APS/123-QED}

\usepackage{hyperref}
\hypersetup{colorlinks=true,linkcolor=blue, filecolor=blue, urlcolor=blue,  citecolor=blue}

\begin{document}

\title{High-Power Laser Drives Motion in Ultra-thin 
Photonic Crystal Lightsails}

\author{Lucas Norder}
\altaffiliation{These authors contributed equally to this work.}
\affiliation{Department of Precision and Microsystems Engineering, Delft University of Technology, Mekelweg 2, 2628CD Delft, The Netherlands}

\author{Ata Keşkekler}
\altaffiliation{These authors contributed equally to this work.}
\affiliation{Department of Precision and Microsystems Engineering, Delft University of Technology, Mekelweg 2, 2628CD Delft, The Netherlands}

\author{Richard A. Norte}
\email{r.a.norte@tudelft.nl}
\affiliation{Department of Precision and Microsystems Engineering, Delft University of Technology, Mekelweg 2, 2628CD Delft, The Netherlands}

\date{\today}

\begin{abstract}
\textcolor{black}{Laser-driven lightsails require reflectors that are ultralight, highly reflective and robust under intense illumination. These demands point to the thinnest reflectors that dielectric optics allows, photonic crystals, in which a perforated membrane thinner than the wavelength of light acts as a mirror. Such mirrors reflect by confining light inside the membrane, but this confinement also amplifies absorption, making ultralow-loss, tension-bearing materials essential. Stoichiometric high-stress silicon nitride uniquely meets both requirements, yet its $\sim$1~GPa stress makes large, compliant structures difficult to release intact, and such sails have never been pushed by high-power lasers. Here we demonstrate millimeter-wide, 200-nm-thick photonic-crystal trampolines in this material, with 99\% design reflectivity at $1070$~nm and a static stiffness of $0.53$~N/m. Radiation pressure displaces them by more than a micrometer within 50~$\mu$s. In air, they withstand several times the irradiance found at the Sun's surface, delivered as a coherent beam. Light reshapes the sail more than it pushes it, since modest temperature differences buckle the membrane and collapse its stiffness, so thermal deformation outpaces radiation pressure within milliseconds. The interaction and compliance that make lightsails movable also let light deform them. These results establish a testbed for high-power nanophotonics, directed-energy systems and light-driven propulsion.}

\end{abstract}

\maketitle

\begin{figure}[ht]
    \centering
    \includegraphics[width=1\linewidth]{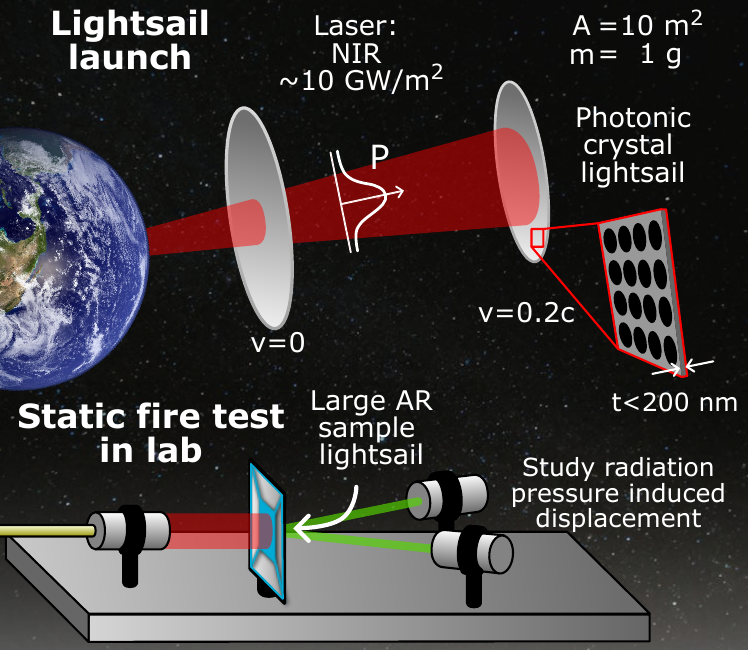}
    \caption{\textbf{Static fire test of lightsail launch in lab setting}. A, area; m, mass; AR, aspect ratio; PhC, photonic crystal; I, intensity of the laser; NIR, near-infrared. In a lightsail mission, a lightsail will be launched from space by being illuminated with a high-power NIR laser. To study the optical behaviour of a lightsail, initial tests must be conducted in the lab. Therefore, large AR lightsail samples are required for uniform illumination, eliminating the need for laser focusing.}
    \label{fig:Fig1}
\end{figure}

Recent advances in high-power lasers, microchip-scale systems, and nanophotonic materials have renewed interest in laser-driven lightsails as a route to accelerating ultralight payloads to extreme velocities \cite{MARX1966,lubin2022,BreakthroughInitiatives} (Fig. \ref{fig:Fig1}). In this concept, a high-power laser beam transfers momentum directly to an ultralight reflective sail, offering a path to accelerate gram-scale spacecraft to substantial fractions of the speed of light, opening a possible route toward interstellar exploration. More broadly, it defines a new regime of extreme light-matter interaction in which large optical powers act directly on ultralight engineered materials.

Realizing this vision requires an unusual form of nanotechnology. Unlike most microchip and nanophotonic devices, which are miniaturized in all three dimensions, a lightsail must maintain nanoscale thickness while extending laterally to much larger scales in order to satisfy strict mass constraints while intercepting large laser beams. This makes lightsails an extreme-aspect-ratio nanophotonic material: a free-standing structure thinner than the wavelength of light, yet spanning macroscopic distances and patterned with millions of nanoscale features, unlike any other large-scale optical component made today. Achieving high reflectivity, low absorption, low areal mass, mechanical robustness, and large-area uniformity within a single structure remains a major challenge.

One particularly promising route is photonic crystal (PhC) based lightsails \cite{lin2025,atwater2018materials}. By patterning dense arrays of wavelength-scale holes into single-layer dielectric membranes, these structures can support resonant optical modes (Fig. \ref{fig:method}a) that produce very high reflectivity despite subwavelength thickness~\cite{fan2002analysis, Moura2018, Lien2022, Stambaugh2015, Norte2016, Enzian2023,Chen2017}. \textcolor{black}{A single layer thinner than the wavelength it reflects, such a membrane is the thinnest dielectric mirror optics allows.} At the same time, the perforations reduce mass, allowing the material to approach the demanding areal densities required for optical propulsion. \textcolor{black}{Because such a photonic crystal reflects by confining light within the membrane, absorption acts precisely where the optical field is strongest, so the material must be chosen for the lowest loss available rather than for ease of fabrication.} \textcolor{black}{High-stress stoichiometric} LPCVD silicon nitride \textcolor{black}{(Si$_3$N$_4$) is the leading candidate, combining sub-ppm} near-infrared \textcolor{black}{absorption}~\cite{Land2024,Kumar2025} with \textcolor{black}{the} tensile mechanical properties~\cite{Wilson2009,Zwickl2008,Xu2024} \textcolor{black}{needed for ultralight, mechanically stable membranes at large scale}~\cite{Norder2025,Moura2018}. \textcolor{black}{The two are inseparable: the stoichiometric films with the lowest absorption are also those carrying $\sim$1~GPa of tensile stress, which makes large membranes stiff, so that structures compliant enough to move measurably under radiation pressure become very difficult to realize.}

At the same time, whether such structures can actually operate under the required optical conditions remains an open experimental question. The same resonant nanophotonic effects that enable strong reflection also imply strong interaction with the incident field, making their robustness under intense illumination far from obvious. Prior work has measured laser radiation pressure on thick macroscopic mirrors \cite{Pinot2019,Wagner2018,Williams2013,Takei2025,Vaskuri2021, Myrabo2002}, which are too heavy for the lightsail application. For single-layer reflectors, only displacements of around 170 picometers have been resolved \cite{Michaeli2025}, roughly the radius of a silicon atom. \textcolor{black}{Such measurements drive the structure periodically and detect the minute response with lock-in techniques, which reject slow thermal deformation by construction. Cavity-based readout, as used in membrane-in-the-middle optomechanics~\cite{thompson2008strong}, is similarly restricted to picometer-scale motion, since it remains linear only over a fraction of a cavity linewidth.} 

Large suspended PhC membranes have been realized~\cite{Moura2018,Norder2025}, but their high intrinsic stress makes them too stiff to generate measurable displacements under available optical power; trampoline geometries dramatically improve compliance in Si$_3$N$_4$~\cite{Rodriguez2011,Kleckner2011,norte2015, Norte2016,Reinhardt2016,Gaertner2018,Jong2022,Afridi2026,Michaeli2025}, yet have never been scaled to the large reflective areas needed for direct lightsail studies. It therefore remains unclear whether one reflector can be large, highly reflective, \textcolor{black}{compliant enough to move measurably under radiation pressure, and robust under intense illumination.}

Here we address this long-standing problem using large-area photonic crystal trampolines: 200-nanometer-thick, millimeter-wide silicon nitride membranes patterned with a photonic crystal and suspended by thin tethers. This nano-engineered architecture combines subwavelength reflectors with the compliance needed to resolve \textcolor{black}{static radiation-pressure deflection under continuous, unmodulated illumination}. Using a high-resolution displacement platform, we measure their response under high-power laser illumination and identify the mechanisms governing their motion (Fig. \ref{fig:Fig1}). Our results show that single-layer large-area PhC membranes can be simultaneously reflective, compliant, and robust under extreme optical loading. Crucially, these experiments reveal dominant thermomechanical effects, including laser-induced buckling \textcolor{black}{at temperature differences of only $\sim$8~$^\circ$C}, that have not previously been observed in this regime and that must be accounted for in any realistic lightsail design.
\section*{Results}

\subsection*{\textbf{Development of photonic crystal lightsail}}

\textcolor{black}{This study probes the forces exerted on a lightsail by high-power laser illumination, both the radiation pressure that underlies lightsail propulsion and the thermomechanical response that accompanies it.} In a Starshot-type scenario, a lightsail must sustain continuous near-infrared illumination at power densities of approximately 10 \si{GW/m^2} for several minutes~\cite{BreakthroughInitiatives}. As a first step toward this regime, we focus here on the deflection of a lightsail material under continuous illumination; a mechanical analogue of a static fire test in rocket development, designed to demonstrate controlled thrust generation under non-accelerating conditions.

\begin{figure*}
    \centering
    \includegraphics[width=1\linewidth]{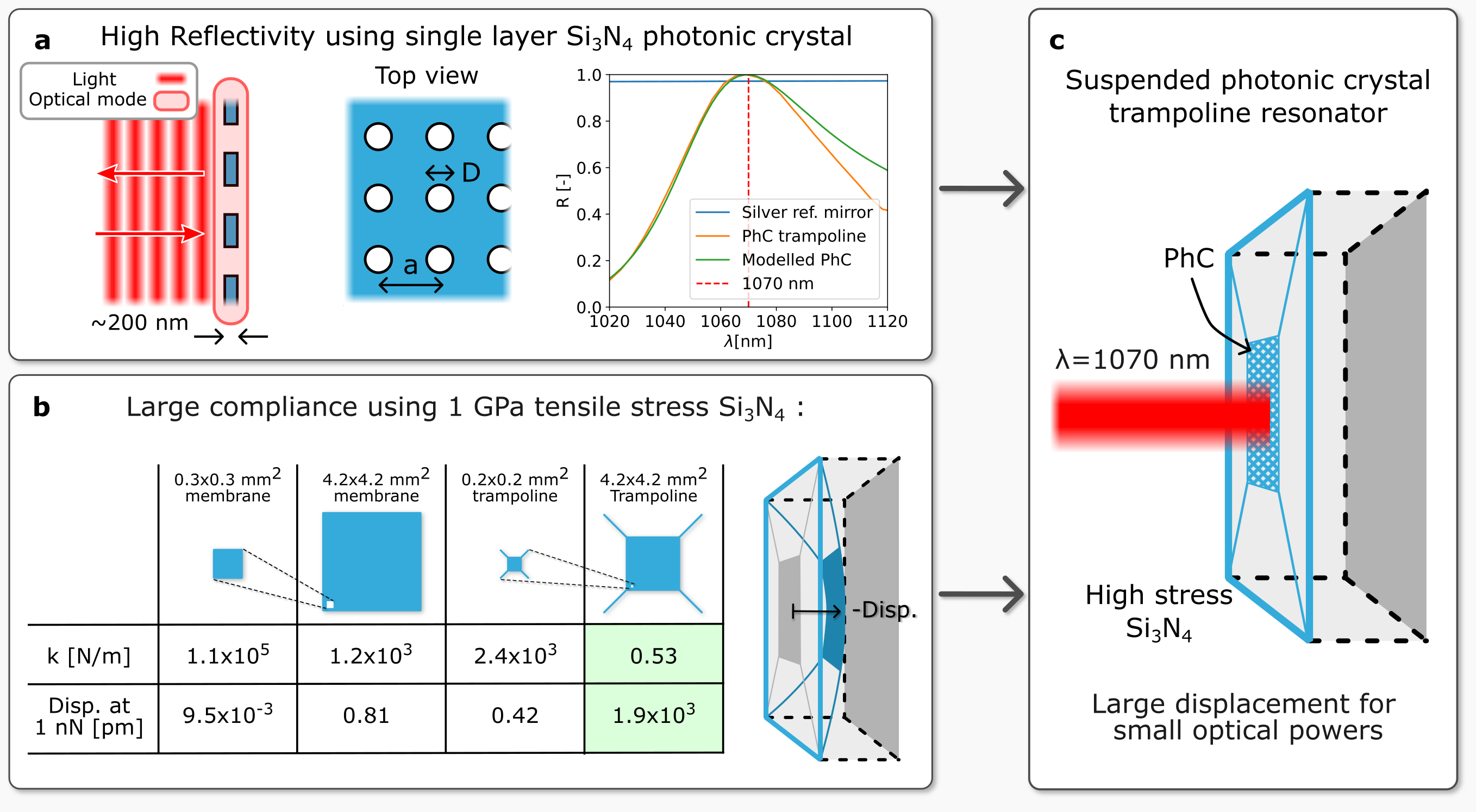}
    \caption{\textbf{Working principle of the photonic crystal trampoline.} a, period; D, diameter; d, displacement; $\lambda$, wavelength of light; Mem, membrane; Tramp, trampoline. \textbf{a}, The working principle of a PhC lightsail, along with the top view and reflectivity spectrum. The spectrum is measured (orange) from a $200$~nm single-layer PhC membrane optimized for $1070$~nm light, with a period of $862$~nm and circular holes of $657$~nm diameter, using a silver mirror (blue) as a reference. The simulation is obtained from an RCWA simulation (green) of an infinite, free-standing periodic membrane, excluding the substrate. The measurement is described in more detail in the Methods section. \textcolor{black}{\textbf{b}, Simulated static stiffness $k$ and displacement under a $1$~nN load for four geometries in 1~GPa tensile \ch{Si3N4}, computed in COMSOL. Scaling the trampoline design to the millimeter scale yields a large compliance despite the high tensile stress of the film.} \textbf{c}, Combining the compliant suspended trampoline with a PhC results in a mechanical system sensitive enough to directly measure its deflection due to radiation pressure.}
    \label{fig:method}
\end{figure*}

To directly resolve radiation-pressure motion, the device must simultaneously satisfy four requirements: high reflectivity, ultralow areal mass, a large illuminated area, and high mechanical compliance. As shown in Fig.~\ref{fig:method}a, high reflectivity at 1070 nm is achieved using a 200 nm-thick high-stress \ch{Si3N4} membrane patterned with a square lattice of holes. The optimized unit cell has a period of 862 nm and circular holes with a diameter of 657 nm, \textcolor{black}{producing a calculated reflectivity of $99\%$, consistent with reflectivities reported for similar single-layer photonic-crystal reflectors \cite{Moura2018}}. A more in-depth reflectivity analysis under operating conditions and polarization dependence can be found in the Supplementary Information. At the same time, the perforated membrane reduces the areal mass while preserving strong resonant interaction with light, making photonic crystal membranes \textcolor{black}{the most promising route to meeting} the stringent requirements of lightsail applications. \textcolor{black}{For an interstellar mission, such sails also require broadband reflectivity to accommodate Doppler shifts during acceleration. The square lattice used here is not optimized for that purpose, and designs offering broader bandwidth have been demonstrated separately~\cite{Norder2025}.} 

\textcolor{black}{The ultra-low optical absorption required for a lightsail is achieved in stoichiometric LPCVD \ch{Si3N4}, the lowest-loss film that can also hold the tension needed to suspend large structures flat, without wrinkling.} This comes with a central challenge: the same high-temperature deposition that produces low defect density leaves the film under about 1~GPa of tensile stress. That stress is beneficial for optical flatness and structural integrity, but it also makes suspended membranes extremely stiff, rendering them insensitive to the small forces generated by radiation pressure.

\textcolor{black}{To overcome this limitation, we adopt the trampoline geometry introduced above, in which the reflective membrane is suspended by thin tethers. Trampolines of this kind originate in cavity optomechanics, where \ch{Si3N4} membranes serve as high-$Q$, optically compatible elements for cavity-enhanced motion sensing. Here we instead scale the geometry to millimeter dimensions and operate it in free space under direct high-power illumination, so that the displacement under the available laser power reaches the micrometer range and can be measured directly~(Fig.~\ref{fig:method}c).}

\textcolor{black}{Fig.~\ref{fig:method}b compares the simulated static stiffness of four geometries in 1~GPa \ch{Si3N4}, drawn to scale from the micrometer-scale pads of previous trampolines~\cite{Jong2022,Michaeli2025,Norte2016} to the millimeter-scale device used here. Increasing the lateral size and introducing tethers each significantly increase compliance: the device used here, the $4.2\times4.2$~mm$^2$ trampoline, reaches a static stiffness of $0.53$~N/m, roughly 200{,}000 times more compliant than the $0.3$~mm baseline \ch{Si3N4} membrane~\cite{Chen2017}. Comparable compliance is routinely reached in soft atomic force microscope cantilevers~\cite{Sader2012}. What is difficult is reaching it at millimeter scale in a 1~GPa film, where the same stoichiometry that gives the lowest optical absorption also brings the tension that resists deflection.}

\begin{figure*}[t]
    \centering
    \includegraphics[width=1\linewidth]{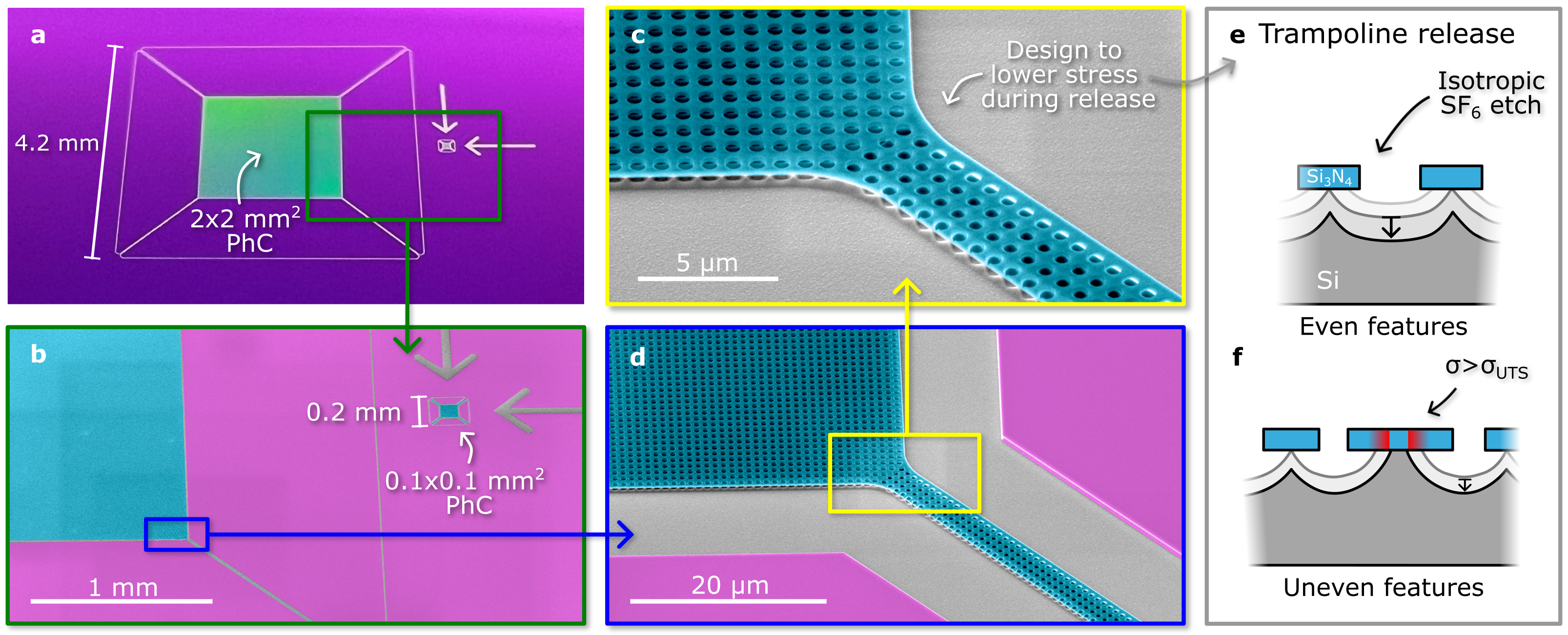}
    \caption{\textbf{Fabricated large-scale PhC trampoline.} \textbf{a}, photograph of PhC trampoline with reference trampoline. \textbf{b}, Coloured SEM image: purple is \ch{Si3N4} on top of Si, blue is suspended \ch{Si3N4}, grey is Si. \textbf{c} and \textbf{d} are close-ups of the corner of the PhC patch. \textbf{e}, Schematic of the \ch{SF6} isotropic etch, releasing the LPCVD high-stress \ch{Si3N4} device. Similarly sized features of \ch{Si3N4} result in equal release rate of the structures with an isotropic \ch{SF6} etch. \textbf{f}, \ch{Si3N4} device with different size features can result in the stress reaching the ultimate tensile strength (UTS) and destroying the sample during the release process.}
    \label{fig:tramp}
\end{figure*}

As shown in Fig.~\ref{fig:tramp}a,b, the final structure consists of a 200 nm-thick suspended \ch{Si3N4} PhC spanning $2\times2$~\si{mm^2}, patterned with millions of nanoscale holes and supported by thin tethers to form a $4.2\times4.2$~\si{mm^2} suspended trampoline. Compared with earlier PhC trampolines~\cite{Jong2022,Michaeli2025,Norte2016}, which were typically confined to micrometer-scale reflective pads, this device extends the reflective trampoline concept by orders of magnitude in lateral scale. \textcolor{black}{Scaling in this way places correspondingly greater demands on the design, since the tethers and the perforation pattern must together carry the full membrane tension while releasing intact across the whole structure.}

Fabricating such a device places stringent demands on nanofabrication. \textcolor{black}{The stress cannot simply be lowered to make fabrication easier. A silicon-rich, low-stress film would release far more readily at these sizes, but its higher optical absorption rules it out for sustained high-power illumination. The 1~GPa stress of low-absorption LPCVD \ch{Si3N4} is therefore fixed, and the device design and release process must be built around it~\cite{Land2024}. The thin tethers that make the trampoline compliant are also where that stress concentrates (Fig.~\ref{fig:tramp}c,d).}

The device was fabricated by depositing high-stress LPCVD \ch{Si3N4} on Si, patterning the PhC by electron-beam lithography, transferring the pattern by plasma etching, and finally releasing the suspended trampoline from the substrate. Rather than using a conventional wet release~\cite{norte2015}, we employ an isotropic \ch{SF6} dry etch, which avoids capillary forces during drying and greatly improves fabrication yield at large areas~\cite{Shin2022, Norder2025}. \textcolor{black}{Reliable release depends on keeping this etch uniform across the entire structure. The isotropic etch rate depends strongly on local feature geometry, so dissimilar features release at different rates, and the resulting non-uniformity generates stress concentrations that can reach the ultimate tensile strength and destroy the device (Fig.~\ref{fig:tramp}e,f). Our solution is to continue the photonic-crystal hole pattern across the whole suspended structure, including the tethers and the irregular clamping regions where the stress is greatest. Projecting a uniform lattice onto these regions, while keeping every hole within the narrow range that preserves the optical resonance, is what lets the membrane release at a single rate and reach the millimeter scale intact.}

\textcolor{black}{To validate the mechanical design, the device was measured using a Polytec laser Doppler vibrometer. The fundamental mode appears at approximately $4.4$~kHz, within $2\%$ of the $4.3$~kHz obtained from the COMSOL eigenfrequency analysis, confirming that the fabricated structure matches the designed mechanical response. Further frequency-domain characterization and an uncertainty analysis are provided in the Supplementary Information.}

\begin{figure*}[t]
    \centering
    \includegraphics[width=1\linewidth]{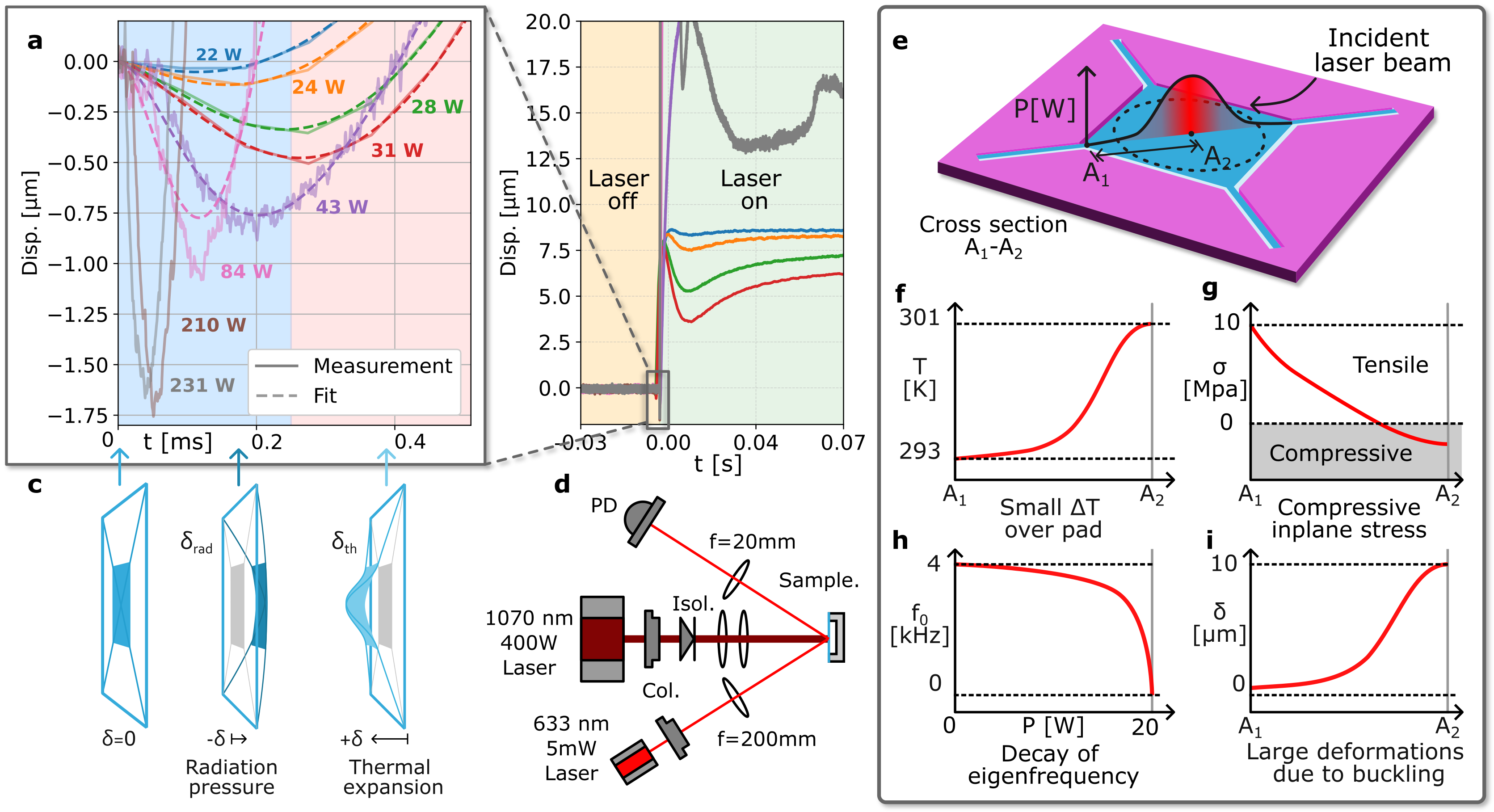}
\caption{\textbf{Radiation-pressure and thermal response of the PhC trampoline under high-power illumination.} $\delta$, displacement; PD, photodiode; Col., collimator; Isol., isolator; f, focal point. \textcolor{black}{\textbf{a}, Zoom-in of the initial response to laser excitation. Negative displacement corresponds to the trampoline moving toward the substrate, in the direction of radiation pressure.} \textbf{b}, Trampoline displacement over the full duration of the measurement. \textbf{c}, Schematic of the trampoline state at different times during the experiment. \textbf{d}, Experimental setup. \textcolor{black}{\textbf{e}, Buckling of the PhC trampoline induced by non-uniform heating from the incident Gaussian laser beam, with the cross section taken along A1--A2. \textbf{f}, Temperature profile due to optothermal heating at an incident power of 20~W. \textbf{g}, Stress profile in the PhC pad at the same incident power. Heating induces compressive stress near the center of the pad, driving buckling. \textbf{h}, Decay of eigenfrequency ($f_0$) with increasing optical power, indicating the onset of buckling. \textbf{i}, Buckling strongly reduces the stiffness of the PhC pad, leading to large deformations. Panels \textbf{f--i} are qualitative trends extracted from FEM analysis underlying the buckling mechanism.}}
    \label{fig:Result}
\end{figure*}

\subsection*{\textbf{Measurements}}
To probe the response of the photonic crystal trampoline under high-power optical loading, we illuminated the device with up to 230 W of optical power from a continuous-wave fiber laser. The beam was linearly polarized and reduced to a 1 mm diameter to match the central PhC region without tight focusing. Resolving the trampoline's motion is experimentally challenging because the measurement must detect micrometer-scale displacements while avoiding placing sensitive optics in the high-power beam path. To meet these constraints, we developed a readout scheme based on confocal microscopy (Fig. \ref{fig:Result}d), described in detail in the Supplementary Information. The readout uses optical lever arms to convert small membrane displacements into measurable shifts in the confocal signal. An important feature of the device geometry is the partially undercut silicon substrate beneath the suspended PhC pad. Because the undercut depth is known with nanoscale precision, contact between the deflecting membrane and the substrate provides an absolute reference for the maximum displacement (see Methods). 

\textcolor{black}{Each measurement illuminates the trampoline at fixed power for 4~s while tracking its displacement, repeated across a range of powers. Two competing effects govern the response. Radiation pressure acts essentially instantaneously and pushes the membrane toward the substrate, while thermal deformation builds over milliseconds ($\tau_\mathrm{h}=4.7\pm0.3$~ms, Supplementary Section~B) and drives it in the opposite direction, so the two are separated by both their timescale and their sign. This is clearly visible in Fig.~\ref{fig:Result}a, which shows the response at the moment the laser is switched on. The trampoline first moves toward the substrate, in the direction of radiation pressure, reaching a maximum displacement of $1.75\pm 0.10~\mu$m at 210~W (Supplementary Section A).}

To quantify these competing contributions, we model the device as a single-degree-of-freedom harmonic oscillator with a time-dependent equilibrium position set by thermal deformation. This model captures the two characteristic time scales and agrees well with the measurements up to 34~W. It then slowly deviates and no longer matches the data above 85~W, indicating the onset of a nonlinear mechanism beyond linear thermal expansion. The model's parameter breakdown, along with a discussion of its limitations, is provided in the Supplementary Information.

At these higher powers, the displacement also begins to saturate as the membrane approaches the silicon substrate. In this regime, near-field coupling between the PhC guided modes and the substrate reduces reflectivity and increases transmission into the silicon, thereby introducing additional thermal effects that the model does not capture. The high-power traces in Fig.~\ref{fig:Result}a are therefore shown without a fit. Importantly, the reflectivity reduction depends on the distance to the substrate rather than on intensity or temperature, as shown by the studies of thermal effects on reflectivity and reflectivity stability during illumination in the Supplementary Information.

\textcolor{black}{A similar distinction between low- and high-power behavior is evident in the trampoline's out-of-plane stability. At lower incident powers, the displacement settles to a stable, power-dependent value. However, at higher powers, the trampoline becomes unstable and no longer settles. As the displacement readout is sensitive to both out-of-plane motion and angular tilt, we cannot attribute this continued motion to either mechanism individually.}

\textcolor{black}{Finite-element analysis in COMSOL provides a qualitative illustration of this mechanism (Fig.~4e). Because the incident Gaussian beam heats the center of the PhC pad more strongly than its edges, the calculated temperature profile (Fig.~4f) shows the pad center reaching a higher temperature than its edges, producing the compressive stress distribution of Fig.~4g, concentrated near the center of the pad. This stress is the origin of the buckling instability: as optical power increases, the simulation shows the eigenfrequency decreasing (Fig.~4h) and the effective stiffness dropping sharply (Fig.~4i), reproducing the qualitative onset of buckling and the associated softening observed experimentally, and leading to the large out-of-plane deformation that strongly amplifies the thermally driven response. We emphasize that these simulations illustrate the mechanism responsible for the nonlinear thermal response rather than quantitatively reproducing it.} Notably, the onset occurs for temperature differences of only about \textcolor{black}{8~$^{\circ}$C}, revealing how modest non-uniform heating can destabilize ultrathin reflective sails under concentrated illumination. This is particularly relevant for lightsail operation, where the launch beam is likewise expected to generate non-uniform optical loading. The present measurements are performed in air, where convective cooling partially mitigates the temperature rise. In vacuum, where this cooling pathway is absent, both the buckling magnitude and the onset power are expected to become substantially more severe. Full details of the dynamical and thermal models are provided in the Supplementary Information.

\textcolor{black}{The measured $1.75$~$\mu$m displacement is roughly $10^{4}$ times larger than previously resolved in single-layer lightsail reflectors~\cite{Michaeli2025}, and large enough to be read out directly from a single unmodulated trace rather than recovered by lock-in detection. This direct readout is what allows the static thermal response to be observed at all. The prompt motion occurs within approximately $0.05$~ms, corresponding to an average acceleration of $2x/t^2 \approx \textcolor{black}{1.4\times10^{3}}$~m/s$^2$, of order $100\,g$, obtained from the displacement trace alone and independent of mass and reflectivity. Uncertainties on both quantities are given in Supplementary Section~\textcolor{black}{A and E}. These displacements are achieved at a power density of $0.3$~GW\,m$^{-2}$, about $3\%$ of that envisioned for a Starshot-type launch, though the illumination duration remains shorter than the multi-minute launch regime.}

\begin{figure}
    \centering
    \includegraphics[width=1\linewidth]{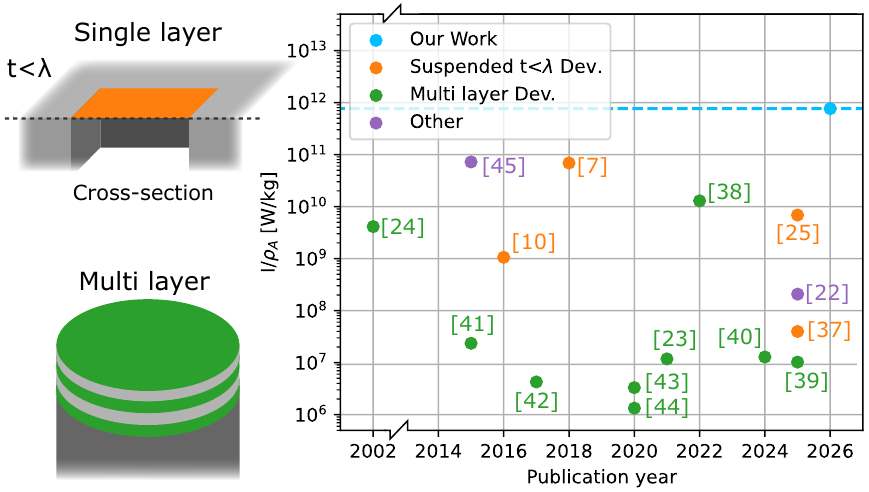}
    \caption{\textbf{Power density over areal density.} I, power density; $\rho_A$, areal density. The state of the art for reflective devices under direct illumination of a high-power-density laser beam.}
    \label{fig:Lit}
\end{figure}

To place these results in context, Fig.~\ref{fig:Lit} compares experimentally tested, directly illuminated reflectors and materials in terms of optical power density and areal density ($\rho_A$)~\cite{Hahtela2005, Moura2018, Yao2025, Norte2016, Michaeli2025, Atikian2022, Luo2025, Li2024, Senger2015, Vaskuri2021, Williams2017, Partanen2020, ArtusioGlimpse2020, Ma2015, Takei2025, Myrabo2002}. \textcolor{black}{The ratio $I/\rho_A$, where $I$ is the single-pass directed power density, compares how much directed optical power a reflector sustains per unit mass, a figure that bears on high-power optics and directed-energy systems as well as on propulsion. It is a materials- and device-level metric rather than a complete lightsail figure of merit, and does not capture absorptance, emissivity, equilibrium temperature or beam-riding stability. The highest intensity reported to date is from a monolithic diamond mirror, 20~GW\,m$^{-2}$ under a 10~kW fiber laser of the class used in directed-energy systems, in one of the best thermal conductors known and clamped to a water-cooled stage~\cite{Atikian2022}. The Si$_3$N$_4$ reflectors here are 1{,}500 times thinner, uncooled, and made from a material that conducts heat some 600 times more poorly, yet reach an $I/\rho_A$ more than an order of magnitude higher.}

High-finesse cavity experiments using \ch{Si3N4} PhC membranes can reach even larger effective $I/\rho_{A}$ values through intracavity power enhancement \cite{Enzian2023,Zhou2023,Chen2017}. However, because cavity power is extremely sensitive to nanometer-scale mirror displacement, these conditions are not directly comparable to directed illumination of a lightsail. Even so, the compatibility of \ch{Si3N4} PhC membranes with high-finesse cavities suggests that this material platform can tolerate still higher optical loading under appropriate conditions.

\begin{figure*}[t]
    \centering
    \includegraphics[width=0.9\linewidth]{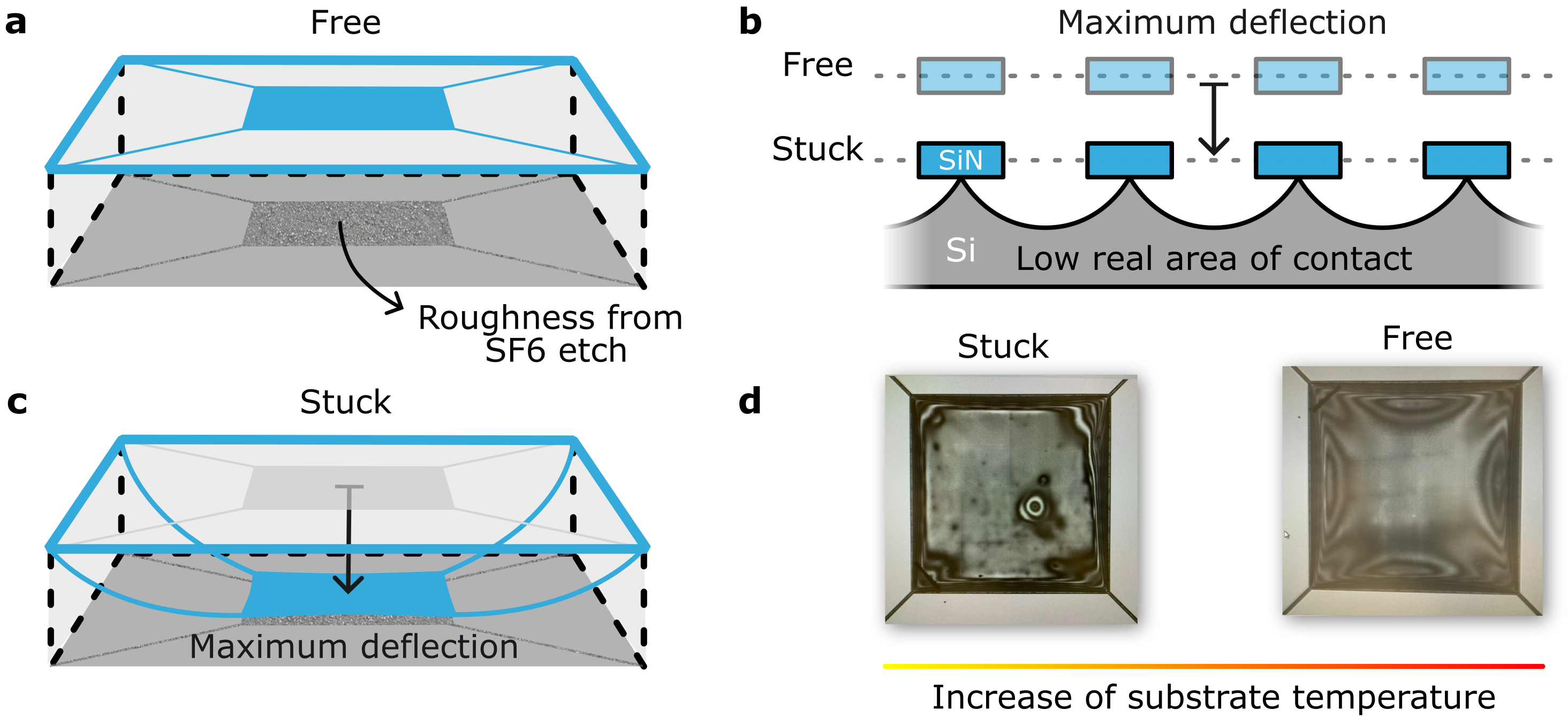}
    \caption{\textbf{Thermal release of a collapsed PhC trampoline.} \textbf{a}, Schematic of free standing trampoline resonator. \textbf{b}, Schematic of contact points of the PhC trampoline and the substrate. \textbf{c}, Schematic of trampoline stuck at the substrate. \textbf{d}, Optical microscopy image of collapsed trampoline and free-standing trampoline after heating the substrate.}
    \label{fig:Stuck}
\end{figure*}

The trampoline proved robust, surviving repeated illumination cycles over a broad range of optical powers. Even contact with the substrate does not damage the device: trampolines that adhere can be released by gently heating the chip, after which the film's tensile stress restores the suspended geometry (Fig.~\ref{fig:Stuck}). The same device can then be measured repeatedly. When failure eventually occurred after extensive testing at incident powers up to 280~\si{W}, the fracture originated in the supporting \ch{Si} substrate rather than in the suspended PhC membrane itself.

\section*{Discussion}
In this work, we have demonstrated \textcolor{black}{unidirectional} radiation-pressure displacement of a large-area ultrathin photonic crystal trampoline under high-power continuous-wave laser illumination. A 4.2$\times$4.2~\si{mm^2} suspended PhC trampoline exhibited a prompt displacement of up to 1.75~$\mu$m under 210~W illumination at 1070~nm, limited by the $\approx2~\mu$m gap to the substrate. \textcolor{black}{The deflection is roughly 10,000 times larger than previously reported optomechanical responses of lightsail candidate materials~\cite{Michaeli2025}, and is sustained under constant illumination rather than recovered from a modulated drive.} This motion occurs within approximately \textcolor{black}{50~$\mu$s}, corresponding to an average acceleration of order 100~\textit{g} driven solely by photon momentum, which the 50~$\mu$s onset separates from the millisecond thermal response. Our group previously showed that single-layer photonic crystal membranes can be scaled to wafer dimensions~\cite{Norder2025}. Here we show that such ultrathin structures \textcolor{black}{also withstand directed laser intensities of 0.3~GW\,m$^{-2}$, roughly five times the 63~MW\,m$^{-2}$ radiated at the surface of the Sun. Thermal viability in flight additionally requires absorptance and radiative balance to be quantified in vacuum.}

At the same time, our measurements show that a lightsail cannot be treated simply as a rigid reflector pushed by light~\cite{lin2025}. Even under modest temperature gradients, non-uniform optical heating produces strong thermomechanical effects, including buckling and a pronounced reduction in stiffness. This behavior is not a \textcolor{black}{barrier}; rather, it reveals an additional design dimension that becomes especially important at large scale. In practical lightsails, thermally induced shape changes are likely to contribute strongly under beam-riding conditions, especially under spatially non-uniform illumination. Notably, even our single-material membrane exhibits significant thermomechanical deformation; in bilayer designs~\cite{Campbell2025,Chang2024}, mismatched thermal expansion coefficients between layers will amplify such effects further, making single-material architectures crucial for high-power laser propulsion~\cite{Atikian2022}. These thermomechanical effects could not have been revealed in cavity-based experiments, where large displacements would detune the cavity and extinguish the intracavity power. Direct free-space illumination at high power is therefore uniquely capable of uncovering the thermomechanical physics that will govern real lightsail operation and these experiments are the first to do so.

This insight also opens up a new opportunity for design. Because these sails are patterned with \textcolor{black}{millions} of nanoscale features, their optical and mechanical responses can be co-designed across the membrane. At smaller lateral scales, engineered variations occur over relatively few unit cells and can introduce scattering and optical loss~\cite{Agrawal2023}. At lightsail scales, by contrast, the pattern can evolve gradually over enormous numbers of unit cells, enabling smooth spatial transitions in reflectivity, phase, stiffness, and thermomechanical response. This suggests a route toward sails whose hole geometry is designed not only for reflection and low mass, but also for controlled deformation, thermal adaptation, and possibly beam-riding stability under high optical loading~\cite{Siegel2019,Gao2024,Campbell2022}.

Beyond lightsail applications, the thermomechanical effects identified here suggest that high-aspect-ratio nanophotonic membranes may find use as light-controlled deformable optical elements, where laser-induced shape changes could be engineered~\cite{zhao2022}. Related reflective structures at smaller scales are already among the most sensitive force sensors in optomechanics~\cite{Norte2016,Enzian2023}, and scaling them upward may open new opportunities for force sensing~\cite{Cupertino2024} and precision accelerometry~\cite{Krause2012,Chowdhury2023,hyatt2026sail}, particularly with high-vacuum operation and cavity-based readout~\cite{Khokhar2026}.

\textcolor{black}{Scaled to a $450$~mm wafer, the largest commercially available format, a sail with the measured areal density and reflectivity of this device has a mass of $54$~mg, and illuminating it entirely at the same irradiance would take roughly $48$~MW. Carrying a $1$~g payload in addition to the sail, five minutes of illumination brings it to about $90$~km/s over an acceleration distance of some $13{,}500$~km, fast enough to reach Mars in $29$~days on laser power alone; the assumptions are given in the Supplementary Information. That power is far beyond what is available today, but the trajectory is steep. Continuous-wave fiber laser output has risen by about four orders of magnitude over the past two decades~\cite{Richardson2010,Jauregui2013,Worden2021}, and $300$~kW-class directed-energy systems have now been demonstrated, with megawatt-class sources under development~\cite{nLIGHT2023}. A sail of this kind does not merely survive intense illumination, it responds to it in ways that can be designed, so that the same light that accelerates a structure can also shape it. As high-power lasers and large-area nanophotonic materials advance together, capabilities that have existed only as concepts move within reach of direct experiment.}

\nocite{*}
\bibliographystyle{apsrev4-1}
\bibliography{main}

\section*{Methods}

\subsection*{Simulation of the photonic crystal trampoline}
The finite element solver in COMSOL is used to model the eigenfrequency response, stiffness, and thermal behavior of the trampoline, using the plate module throughout.

The trampoline membrane is made from \ch{Si3N4} and is patterned with a PhC hole array across its entire surface. Explicitly resolving all holes in the finite element model would be computationally expensive and not practical. Therefore, an effective material property is implemented. The holes reduce both the effective mass and stiffness of the membrane, each handled separately. The reduced mass is modeled by multiplying \ch{Si3N4} density by the volume fraction, which is defined as the fraction of material remaining after patterning. This is applied to both the pad and the tethers. The reduced stiffness requires a different treatment, since the tethers carry the dominant mechanical load. An analysis of the tether loading shows it is almost purely tensile, making bending stiffness negligible. A small representative tether section with holes is therefore simulated in COMSOL under tensile loading, yielding an effective stiffness of 30\% of the nominal value; the Young's modulus is scaled accordingly. These effective material properties are used across all three simulation studies described below.

For the eigenfrequency analysis, a stationary step first applies the 1~GPa tensile pre-stress intrinsic to the LPCVD stoichiometric \ch{Si3N4} deposition, after which the eigenfrequency analysis is performed. For the stiffness simulation, a Gaussian area load is applied to the membrane to replicate the laser beam profile, and the resulting displacement is used to extract the effective spring constant. For the thermal simulation, three sequential stationary steps are used: first the tensile pre-stress is applied, then the radiation pressure, and finally a Gaussian heat source modeling laser absorption, yielding the membrane's final thermomechanical displacement. For more details on the COMSOL analysis see the Supplementary Information.

\subsection*{Nanofabrication of the photonic crystal trampoline}
For the fabrication of the suspended PhC lightsail, a similar approach is used as described by Shin et al. (2022) \cite{Shin2022}. Initially, a 100 mm Si wafer is covered with 200 nm of \ch{Si3N4} using low-pressure chemical vapour deposition (LPCVD) to attain a pre-stress of 1 \si{GPa}. Next, the wafer is diced into $10\times10$ \si{mm^2} chips, which are coated with ARP-13 positive-tone e-beam resist. The resist is then spin coated at 2000 rpm to reach 550~nm thickness and baked at 155 \si{\degreeCelsius} for 3 min. A Raith EBPG-5200 electron beam writer is used to expose the resist with a 380 \si{\mu C/cm^2} dose. The development consists of a pentyl-acetate for 60 s, followed by MIBK and IPA in a 1 to 1 solution for 60 s. The resist mask enables the PhC pattern to be transferred using a reactive ion etcher (RIE) for 1200 s to etch the \ch{Si3N4} layer with \ch{CHF3}. Finally, a 15 s fluorine-based \ch{SF6} inductively coupled plasma RIE (ICP-RIE) etch is used at -120 \si{\degreeCelsius} to suspend the PhC membrane. 

\subsection*{ \textcolor{black}{Reflectance spectroscopy of the photonic crystal}}

The reflectance spectrum of the fabricated PhC trampoline (Fig.~2a, orange curve) was measured at normal incidence using an Avantes AvaLight-DH-S unpolarised white-light source coupled to an AvaSpec-NIR256 spectrometer. A dark spectrum was first acquired by blocking the light path to the spectrometer and subtracted as background from all subsequent measurements. The probe beam had a diameter of $1$~mm and was incident at the center of the $2\times2$~mm$^2$ PhC pad. The measured spectrum was referenced against a silver mirror to obtain absolute reflectance.

For comparison, the reflectance of the same unit cell was calculated using rigorous coupled-wave analysis (RCWA) and is shown as the green curve in Fig.~2a. This calculation assumes an infinite, free-standing periodic membrane and does not include the silicon substrate beneath the fabricated device. This measurement, therefore, serves two purposes: it confirms that the resonance peak of the fabricated device falls where designed, and it validates the free-standing RCWA model against the measured device, specifically near resonance, where reflectivity is high, and the guided-mode response dominates over any near-field interaction with the substrate beneath. Away from resonance, where reflectivity is lower, the comparison to a free-standing model is less representative of the suspended device. Reflectivity of the full trampoline under the illumination conditions used in the high-power measurements is instead obtained separately and in situ, via the isolator pickoff described in the Supplementary Information (Fig.~S10c).

\subsection*{Unsticking the photonic crystal trampoline}
The isotropic \ch{SF6} etch used to suspend the trampoline produces two characteristics that are both advantageous for device operation. First, the etch leaves a rough surface beneath the freestanding trampoline (Fig. \ref{fig:Stuck}b). This reduces the actual contact area of the trampoline when it is collapsed onto the substrate under excessive load (Fig. \ref{fig:Stuck}a,b) and therefore improves recovery. For recovery, the chip is placed on a temperature-controlled stage and slowly heated to 80 $^{\circ}$C. This is sufficient to fully release the membrane, as confirmed optically (Fig. \ref{fig:Stuck}d). Second, the isotropic etch creates a well-defined undercut depth beneath the trampoline. This sets a hard mechanical limit on the out-of-plane deflection, and the fully collapsed state therefore serves as a convenient in-situ reference for calibrating absolute displacement.

\subsection*{Measurement setup for deflecting by radiation pressure}
The deflection of the PhC trampoline is measured in the schematic setup shown in Fig. \ref{fig:Result}d. The setup consists of both an actuation and a measurement component. An IPG Photonics YLR-400-LP-AC-Y11 laser operating at a wavelength of 1070 nm is used for actuation. This laser beam passes through an isolator and two lenses that can adjust the beam diameter. A secondary 5 \si{mW} and 633~\si{nm} wavelength laser is used for detecting the actual deflection. The probe has a $1/e^2$ diameter of $\approx0.42$~mm on the trampoline; arriving at $60^\circ$ from the normal, its footprint is elongated to $\approx0.42\times0.84$~mm$^2$, within the $2\times2$~mm$^2$ pad and smaller than the 1~mm actuation beam (Supplementary Table~S1). This low-power laser first passes through a lens that focuses the beam onto the PhC trampoline, then through a second lens that collimates it before it reaches the photodiode (PD). This method employs the same principle as confocal microscopes, utilizing a long optical lever to achieve sub-micrometer resolution. For calibration, a Precision Instruments PI-M-230 stage is used to determine the positions at which the PhC trampoline is in and out of focus of the laser, resulting in a zero and a high measurement signal, respectively. The measurement is then taken at the highest PD signal-to-position ratio, which identifies the most sensitive measurement region. More details of the measurement setup are discussed in the Supplementary Information.

\section*{Acknowledgements}

We want to thank Peter Steeneken,  Ruben Guis, Paulina Castro Rodríguez, Mark Kalsbeek, R. Tufan Erdogan, Zichao Li, Tanuj Kumar, and Sonny White for stimulating discussions. Furthermore, we want to thank Francesco Stallone and Hande Aydogmus for their help in the device fabrication. Funded/Co-funded by the European Union (ERC, EARS, 101042855). Views and opinions expressed are however those of the author(s) only and do not necessarily reflect those of the European Union or the European Research Council. Neither the European Union nor the granting authority can be held responsible for them. R.N. would like to acknowledge support from the Limitless Space Institute’s I$^2$ Grant and from the project, Probing the physics of exotic superconductors with microchip Casimir experiments (740.018.020) of the research programme NWO Start-up which is partly financed by the Dutch Research Council (NWO).

\clearpage

\foreach \x in {1,...,25}
{%
\clearpage
\includepdf[pages={\x}]{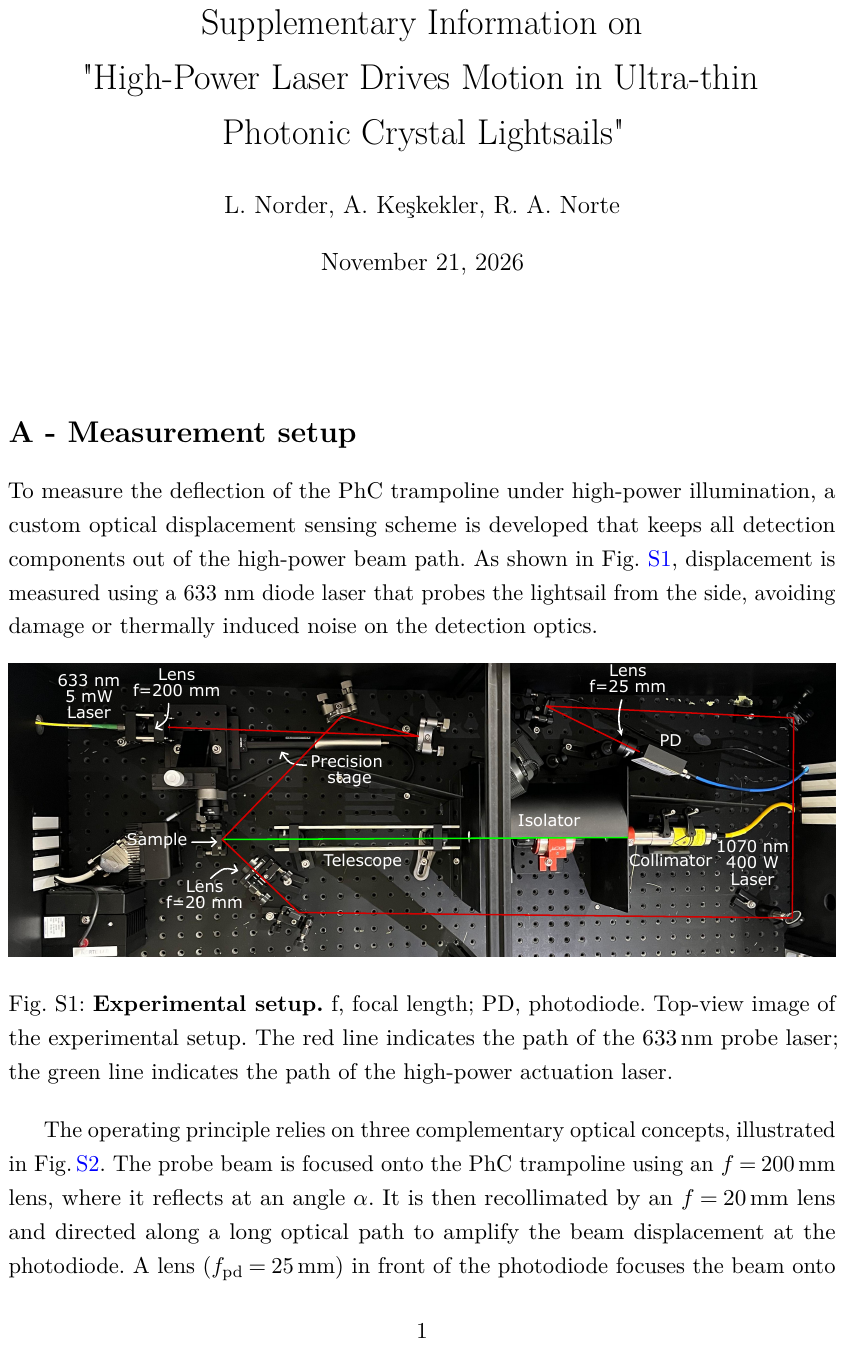} 
}

\end{document}